# A Polarization-insensitive and High-speed Electro-optic Switch Based on a Hybrid Silicon and Lithium Niobate Platform

SHENGQIAN GAO,[1] MENGYUE XU,[1] MINGBO HE,[1] BIN CHEN,[2] XIAN ZHANG,[2] ZHAOHUI LI,[1] LIFENG CHEN,[1] YANNONG LUO,[3] LIU LIU,[2] SIYUAN YU,[1,4] AND XINLUN CAI[1,*]

[1] *State Key Laboratory of Optoelectronic Materials and Technologies and School of Electronics and Information Technology, Sun Yat-sen University, Guangzhou 510000, China*
[2] *South China Academy of Advanced Optoelectronics, South China Normal University, Guangzhou 510000, China*
[3] *Guangxi Key Laboratory of Thalassemia Research, Guangxi Medical University, Nanning, 530021, China*
[4] *Photonics Group, Merchant Venturers School of Engineering, University of Bristol, Bristol BS8 1UB, UK.*
* *chenlf37@mail.sysu.edu.cn, liu.liu@coer-scnu.org, caixlun5@mail.sysu.edu.cn*

**We propose and demonstrate a polarization-insensitive and high speed optical switch unit based on a silicon and lithium niobate hybrid integration platform. The presented device exhibits a sub nano-second switching time, low drive voltages of 4.97 V, and low power dissipation due to electrostatic operation. The measured polarization dependence loss was lower than 0.8 dB. The demonstrated optical switch could provide as a building block for polarization-insensitive and high-speed optical matrix switches.**

Fast optical switch (FOS), with switching times in the micro- to nano-second region, is one of the key enabling optical components for optical packet switching (OPS) [1] and optical burst switching [2]. It brings the benefits of bit rate and format transparency which provide greater agility and flexibility to optical networks. Recently, the rise of high-performance computing and Data Centre Networks (DCNs) in recent years has created a need for FOS, which could enable high bandwidth, low latency, energy-efficient optical interconnect among the servers and racks [3-5]. Leveraging an advanced complementary metal-oxide-semiconductor (CMOS) manufacturing process, silicon photonics has emerged as a powerful platform for high-density photonic integrated circuits to the possibility of low-cost and high-volume production of photonic integrated circuits (PICs) [6-10]. In the past few years, silicon photonic switches have been reported exploiting the thermo-optic (TO) effect [11-15], the free-carrier dispersion effect [16-21], and micro-electro-mechanical-systems (MEMS) technology [22-24]. TO switches suffer from slow switching speed in the order of tens of microseconds or even milliseconds. To achieve nanosecond-scale switching times, free-carrier dispersion effect through carrier injection or depletion is widely exploited for high-speed electro-optic (EO) silicon switch fabrics. Unfortunately, free-carrier dispersion is intrinsically absorptive, degrading not only the insertion loss but also the extinction ratio of the switches. MEMS-actuated optical switch fabrics exhibit low insertion loss, excellent extinction ratio and large port number, but require a high turn-on voltage of >50 V, which complicates the driver design and limits its application. To date, silicon photonic switches with high switching speeds, high extinction ratio and low drive voltage remains a challenging research objective.

Previously, we have demonstrated ultra-high speed and low loss Mach–Zehnder (MZ) modulators based on hybrid integration of lithium niobate (LN) phase shifter with passive silicon circuitry [25]. The devices exhibited low insertion loss of <2.5 dB, a modulation efficiency of 2.2 Vcm, and EO bandwidth of more than 70 GHz. In this paper, we demonstrate a polarization-insensitive MZ switch based on hybrid integration of LN phase shifter with silicon photonic circuits. The presented devices show sub-nanosecond switching speed, low drive voltages of around 5.97 V and low polarization dependence of <0.8 dB. Moreover, the present devices feature energy-efficient electrostatic operation with no power dissipation when holding the switch in the cross or bar state.

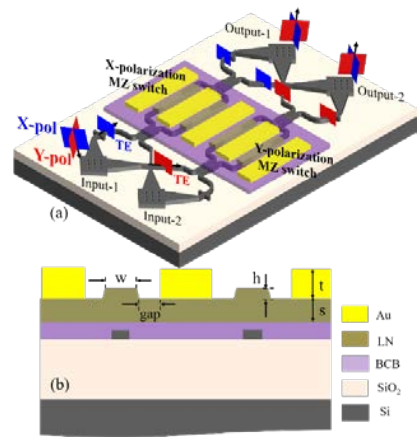

**Fig. 1.** (a) Schematic structure of the polarization insensitive FOS unit; (b) Cross section view of the X(Y)-polarization MZ switches.

A schematic diagram of the polarization insensitive FOS unit is shown in Fig. 1(a). The device consists of a bottom silicon waveguide layer, a top LN waveguide layer and vertical adiabatic couplers (VACs) which transfer the optical power between the two layers. The top waveguides, formed by dry-etching of an X-cut LN thin film, serve as high-speed EO phase shifters where ultra-fast Pockels effect occurs. The bottom silicon circuit supports all of the passive functions, consisting of 3dB multimode interference (MMI) couplers that split and combine the optical power, and two-dimensional grating couplers (2D-GC) for polarization-insensitive off-chip coupling. The VACs, formed by silicon inverse tapers and superimposed LN waveguides, serve as interfaces to couple light up and down between the silicon waveguides and LN waveguides. A mode calculation result (using finite difference eigenmode solver, Lumerical Mode Solution [26]) indicates that nearly 100% optical power can be transferred from silicon waveguide to the LN waveguide, and vice versa [25].

The input signal, coupled through the 2D-GC, is decomposed into two orthogonal polarization components, X-polarization (X-pol) and Y-polarization (Y-pol), and are coupled into a pair of orthogonal waveguides, both in the TE mode (see in Fig. 1(a)). Sharing the same polarization and dispersion, they are switched by the corresponding X- and Y-polarization MZ switches designed for only TE mode, and then sent to the two output 2D-GC (Output-1 and Output-2). The Cross section the device is shown in Fig. 1(b). The LN waveguides have a top width of w=1μm, a slab thickness of s=420nm, a rib height of h=180nm. The thickness of electrodes was set to t=600nm, and the gap between the waveguides and electrodes was set to 2.75μm. The electrodes are designed in a single-drive push–pull configuration, so that applied voltage induces a positive phase shift in one arm and a negative phase shift in the other. The length of the arms of the MZ switches are designed to be 4 mm.

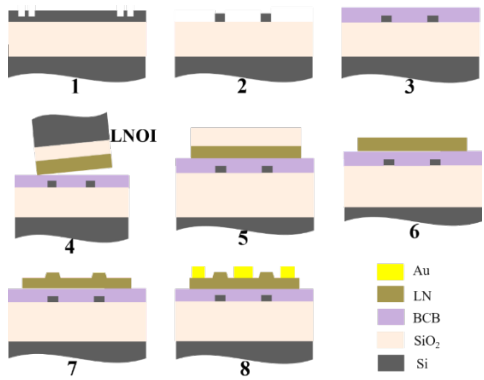

**Fig. 2.** Fabrication flow of the proposed polarization insensitive FOS unit.

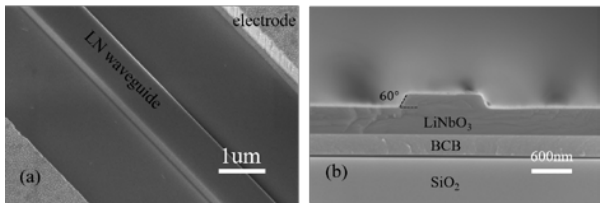

**Fig. 3** SEM image of (a) LN waveguide, and (b) cross section of X(Y)-polarization MZ switches.

The device fabrication process is shown in Fig. 2. The device was fabricated in a silicon-on-insulator (SOI) wafer with 3-μm thickness buried oxide (BOX) and 220-nm thickness silicon waveguide. Firstly, a shallow etched 70 nm 2D-GCs and a 220nm Si waveguide were defined by e-beam lithography (EBL) and inductively coupled plasma (ICP) using hydrogen bromide (HBr), successively. Then a X-cut LN on insulator (LNOI) wafer with silicon substrate, commercially available from NANOLN, was filp-bonded to the patterned SOI wafer through an adhesive bonding process using benzocyclobutene (BCB). After that, the substrate of the LNOI was removed by mechanical grinding and ICP. Then, the BOX layer was removed by a dry etching process. Hydrogen silsesquioxane (HSQ, FOX-16 by Dow Chemical) was then spin-coated on the 600-nm thick LN membrane followed by EBL patterning. Through plasma etching in an inductively coupled (ICP) etching system, the waveguide patterns are transferred into LN. Finally, a liftoff process was performed to produce the Au electrodes. The scanning electron microscope (SEM) image of the fabricated electrode and LN waveguide are shown in Fig. 3(a). Fig. 3(b) shows the cross-section of the fabricated LN waveguides with a sidewall angle of 60°. The total footprint of the device is about 6.0 mm×1 mm.

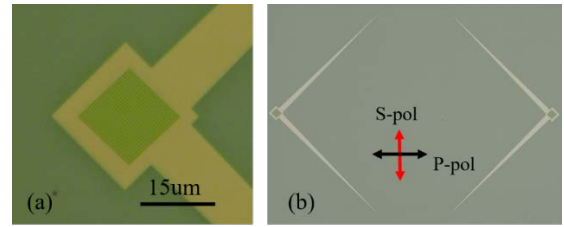

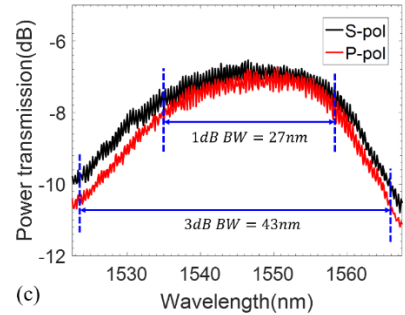

**Fig. 4.** (a) Optical micrograph of 2D-GCs; (b) Optical micrograph of the devices in back to back configuration; (c) The measured P- and S-coupling spectra of the proposed 2D-GCs.

The 2D-GCs are the key components for realizing the polarization insensitive operation [27-28]. The optical micrograph of the 2D-GCs used in the present device is shown in Fig. 4 (a). To measure the PDL, one of the most important performance metrics, two identical 2D-GCs were connected in a back-to-back configuration as shown in Fig. 4 (b). The measured coupling spectra for P- and S- polarization, illustrated in Fig.4 (c), indicate that the PDL is less than 0.8 dB for C-band. The P- or S-polarized input light was calibrated by measuring the transmission of a TE grating coupler for S-polarization, which was co-fabricated with the present device. As shown is Fig. 5, the measured coupling efficiencies is -6.9 dB at the central wavelength of 1547 nm and the 1-dB and 3-dB bandwidth are measured to be 27 nm and 43nm, respectively. The coupling efficiency of the present 2D-GCs is relatively low due to the unoptimized thickness of the BOX layer in the current SOI wafer (3 μm). The coupling efficiency can be significantly improved by using a substrate transfer technique as demonstrated in ref. [28].

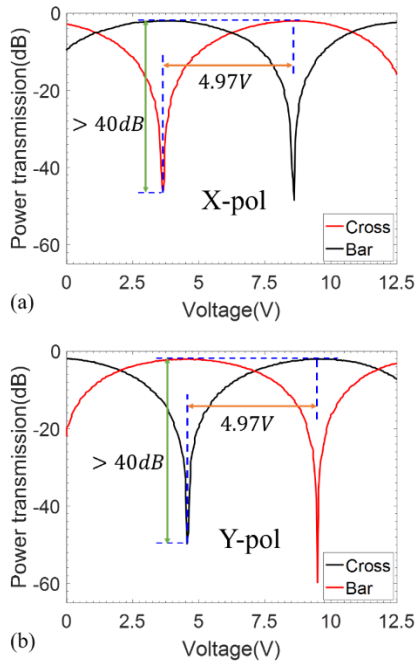

**Fig. 5.** The measured transmission of (a) X-, and (b) Y-polarization MZ switches at different driving voltages.

The measured transmission of the X- and Y-polarization MZ switches at different driving voltages and a fixed wavelength of 1550 nm are shown in Fig. 5. Light wave from a wavelength tunable laser was coupled to the waveguides of the device via a polarization controller (PC) and a single mode fiber. Several TE grating couplers were co-fabricated on the chip in order to calibrate the input polarization states to be X-, or Y-polarization. The transmittances at the Output-1 and Output-2 ports, when the X- or Y- polarization was introduced to Input-1 port, are plotted in Fig. 5 as a function of the voltages applied to the electrode. The X-polarization switch takes the cross/bar state at a voltage of 8.59 V/3.62 V, while Y-polarization switch takes the cross/bar state at a voltage of 9.52 V/4.55 V. The measured Dc $V_\pi$ is 4.97 V. Thus, the circuit would give a polarization insensitive cross or bar state when both of the X- and Y-polarization switches take their cross or bar state. Both of the extinction ratio of the switches for X- and Y-polarization was measured to be > 40 dB, as shown in Fig. 5. The on-chip insertion loss of the polarization diversity switch was estimated to be around 2 dB by subtracting the coupling loss of 2D-GCs. It should be noted that the circuit is very energy efficient because it consumes power only when the state changes, and no power is consumed when holding the switch in the cross/bar state.

To examine the spectral response of the circuit, the transmittance spectra of the cross and bar ports in the cross and bar states for X-, Y-, and a mixed polarization are shown in Fig. 6. Extinction ratios of 26 dB and 28 dB have been achieved in the C-band at both cross and bar output ports for X- and Y-polarizations, respectively. In addition, Fig. 6(c) shows the transmission spectra for the S-polarization, which forms an angle of approximately 45 degrees to the X- or Y-polarization directions. An extinction ratio of at least 28 dB was obtained for S-polarization, which further confirms the broadband polarization insensitive operation of the circuit.

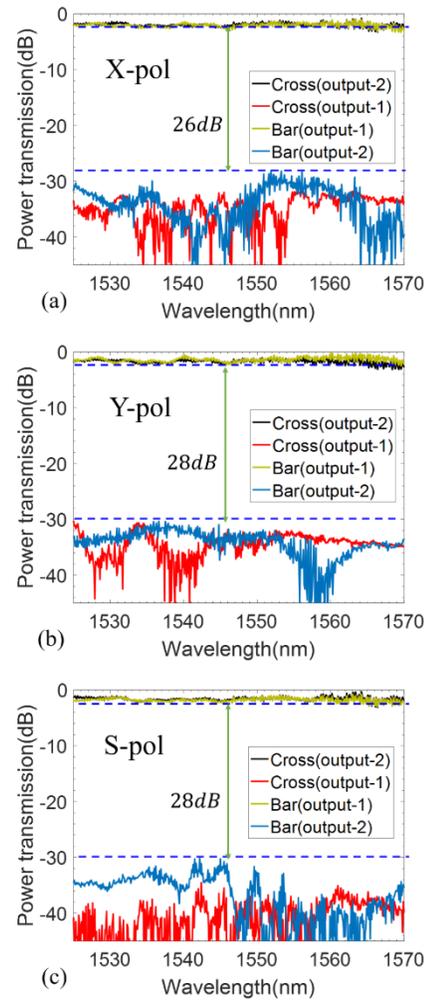

**Fig. 6.** The transmittance spectra of the cross and bar ports in the cross and bar states for (a) X-, (b) Y-, and (c) S-polarization.

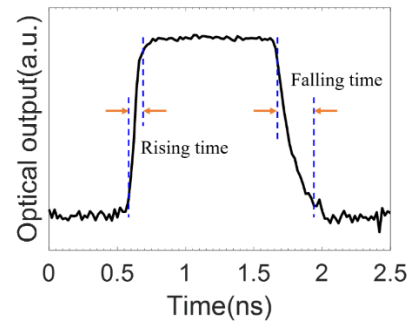

**Fig. 7.** Temporal response of the EO switches.

Finally, we characterized the dynamic switching properties of the EO switches. A square-wave electrical signal with a repetition rate of 500MHz and a duty cycle of 50%, generated from an arbitrary signal generator (MICRAM), was applied to the electrode of the phase shifter through a RF probe. A 50GHz broadband amplifier (SHF 807) was used to amplify the driving signal to the switch together with a DC bias. The optical output intensity is recorded using an oscilloscope (Tektronix DSA8300). As shown is Fig. 7, the rising and falling times were measured to be 100 ps and 312 ps, respectively, indicating an ultra-fast switching speed.

In conclusion, we have designed and demonstrated a polarization-insensitive and high-speed optical switch circuit based on the hybrid silicon and LN platform. The polarization-insensitive operation was achieved with a polarization-diversity technique by using 2D-GC with low PDL of less than 0.8 dB in C-band. The demonstrated device exhibits switching speed of less than 1 ns, an insertion loss of 2 dB and a low drive voltage of around 4.97 V. The switch demonstrated here could provide as a building block for polarization-insensitive, high-speed and large-scale silicon photonic matrix switches.

**Funding.** National Natural Science Foundation of China (NSFC) (11690031, 61575224, 61622510, 61675069), Local Innovative and Research Teams Project of Guangdong Pearl River Talents Program (2017BT01X121), Guangzhou Science and Technology Program (201707010444)